# Computer-Mediated Consent to Sex: The Context of Tinder

DOUGLAS ZYTKO, Oakland University, USA
NICHOLAS FURLO, Oakland University, USA
BAILEY CARLIN, Oakland University, USA
MATTHEW ARCHER, Oakland University, USA

This paper reports an interview study about how consent to sexual activity is computer-mediated. The study's context of online dating is chosen due to the prevalence of sexual violence, or nonconsensual sexual activity, that is associated with dating app-use. Participants (n=19) represent a range of gender identities and sexual orientations, and predominantly used the dating app Tinder. Findings reveal two computer-mediated consent processes: consent signaling and affirmative consent. With consent signaling, users employed Tinder's interface to infer and imply agreement to sex without any explicit confirmation before making sexual advances in-person. With affirmative consent, users employed the interface to establish patterns of overt discourse around sex and consent across online and offline modalities. The paper elucidates shortcomings of both computer-mediated consent processes that leave users susceptible to sexual violence and envisions dating apps as potential sexual violence prevention solutions if deliberately designed to mediate consent exchange.

## 1 INTRODUCTION

Once a taboo topic, sex has gained in popularity as an explicit research focus in human computer interaction over the last two decades [18,20,60,107]. Much of this research has been into sexual wellness, or how technologies can positively augment our sexual lives [8,35,59,100,117]. Other research has focused on sexual violence—meaning activity of a sexual nature without consent [11]—including how the use of technology enables sexual violence, particularly online sexual harassment [16,53,72,87,109], and how technologies could be designed to address this behavior online and offline [1,4,17,68,80,88].

A central characteristic of sexual violence (SV) is the absence of consent, by which we refer to unambiguous agreement to a sexual experience [10]. Despite ample HCI research into SV, we lack understanding of how consent itself is computer-mediated. How does computer-mediated communication shape the ways we give and (perceive to) receive consent to activities of a sexual nature? This is a pressing question given the increasing use of computer-mediated tools for finding and interacting with sexual partners, and mounting calls for empirical investigation into how technology facilitates SV [46,53].

An ideal focal point for studying computer-mediated consent would be dating apps, not only because they are popular tools for discovering sexual partners [6], but because the literature portrays dating apps as unique facilitators of SV across virtual *and* physical modalities [25,45,46,78,86,92,108]. Dating apps are a class of mobile social matching systems, or systems that recommend people to people [103]. Popular examples include Tinder, Bumble, Grindr, and OkCupid. Quantitative research has linked dating app-use with physical-contact SV including rape and bodily injury [25,45,86,108], as well as online SV [46,78]. Yet, as noted by Gillett [46], understanding of how or why dating apps facilitate SV "primarily stems from anecdotal accounts in popular and social media" (p. 217). For example, popular media has brought attention to user struggles with reporting SV to dating app companies [29,55,70].

It is paramount that the role of online dating interfaces in mediating consent exchange be empirically investigated so as to inform interface designs that can curb nonconsensual experience. Given that dating apps have been used by almost half of adults under 30 years of age [6], insight into how interfaces scaffold consent could be the basis for future dating app designs that serve as scalable preventive solutions to sexual violence, which is currently a serious public health issue [24,30,95,96].

In this paper we report a semi-structured interview study (n=19) with dating app users representing a range of gender identities and sexual orientations about their computer-mediated consent processes, primarily through use of the app Tinder. Empirical contributions include:

*Identification of two computer-mediated consent processes:* 1) consent signaling, in which the dating app interface is used to infer and imply consent to sex without any explicit validation of consent prior to sexual activity, and 2) affirmative consent, in which the dating app interface is used to establish a pattern of overt dialogue about sex across online and offline modalities.

*Elucidation of challenges to both of these computer-mediated consent processes* that leave users susceptible to sexual violence, such as misinterpretation of inferred consent, and changing one's mind about sex that was previously agreed to over computer-mediated communication.

## 2 BACKGROUND

In this section we first review HCI research into sex and sexual harm, which we use to argue for more research into computer mediation of consent to sex. The concept of consent to sex, including modern day consent exchange practice, is discussed. We then identify and explore online dating as a context for studying and designing for computer-mediated consent.

### 2.1 HCI and Sex

Human computer interaction research foregrounding sex has stretched back over 15 years to when Blythe and Jones recognized pornography as an oft unspoken, yet dominant, driver of Internet use [18]. This has been followed by research agendas for explicitly examining sex and sexuality in HCI [20,59,60], with some literature mapping the subject area to human rights and social justice [8,9]. Research in this area has commonly explored how current and future technologies may augment sexual gratification and wellness, particularly with sex toys [8,35] and sex robots [97,100,101,107].

Another focus has been on the role of technology in sexual violence (SV)—a term used by the CDC to comprise any activity of a sexual nature without the consent of all parties involved, ranging from sexual harassment to rape [11]. Online sexual harassment through social media platforms like Facebook, Twitter, and even social VR has been shown to be commonplace [16,57,74,87,109]. This includes behavior such as unsolicited sexual comments and the sending of sexually explicit pictures without consent of the recipient [72].

Research has also studied and proposed technology to stop, or support victims of, SV in the physical world. Much of this research has focused on sexual harassment and assault of women in urban areas [17,89,90] such as Bangladesh [1,4,67]. That research has led to prototypical tools to support women in mitigating risk of SV as they navigate their urban environment, such as with panic buttons to alert trusted contacts [1,88,89], user-generated reports [1,31], and safe routes [4,88]. Other research has focused more specifically on rape with systems that match victims to lawyers [80], and stick-on clothing sensors that use auditory alarms and odor-emitting capsules to ward off attackers [68]. The use of social media to solicit support after sexual abuse and other sexual experiences has also been investigated [5,81].

## 2.2 Computer-Mediated Consent to Sex

While nonconsensual sexual experience has been a recurrent focus in the HCI literature, there has been little investigation of how computer-mediated communication shapes understanding of consent itself—how it is given and how it is (perceived to be) received prior to a sexual act such as penetrative intercourse. By consent we refer to the CDC's definition: "words or overt actions [...] indicating a freely given agreement to have sexual intercourse or sexual contact" [11].

Study of computer-mediated consent is needed because SV is a serious public health issue [24,30,95,96], and technology that mediates consent could be a scalable preventative solution. SV is also a gendered issue, hence technologies that effectively mediate consent would then also be impactful to reducing gender inequality. While all genders are impacted by SV, women are disproportionately victimized compared to men [96], as are LGBTQIA+ individuals compared to cisgender heterosexual individuals [19]. For example, over one third of US women, and 35% of women worldwide [42], have experienced SV involving physical contact, compared to a quarter of US men [96]. Over half of transgender men (51%) and more than one third of transgender women (37%) in the United States have experienced sexual assault in their lifetime [56]. Approximately half of non-binary individuals have experienced sexual assault [56].

To lend necessary context to computer-mediated consent, we must first consider how consent is exchanged during face-to-communication; in other words: the preexisting consent practices that computers may now augment. The specific behaviors qualifying as consent to sex during face-to-face communication have been points of contention socio-politically over several decades [69,73]. Best practices for consent exchange as advocated in literature, law, and by sexual health organizations require mutual, overt agreement from all partners [23,58,91,124,125]. This is sometimes called affirmative consent. The onus is on the initiator to receive agreement before engaging in a sexual act, rather than on the recipient to overtly refuse. The absence of overt resistance to sex or escalation in sexual activities (i.e., silence) is not considered consent.

Affirmative consent, despite its potential to mitigate inadvertent SV, is not universally adopted in physical world contexts. In reality, consent practices vary greatly and are often susceptible to SV. For example, non-verbal cues are commonly interpreted as consent to sex [58,69,73]. This is problematic because such cues can be misinterpreted, and individuals sometimes do not overtly stop a sexual act that they otherwise do not want due to fears of retaliation including physical violence and repercussions to their career and social standing [52,62].

In-person consent practices are informed by sexual scripts [94], or messages learned through cultural or social transmission about how to recognize, and behave during, sexual encounters [40]. Many sexual scripts are instigative of SV. Some are based on spatial and temporal factors [22,54], for example: inviting someone to a college dorm room late at night may imply that sex is going to happen, which makes college students uncertain of their sexual agency (their right to deny a sexual advance [84]) in this situation [54, p. 8]. SV victims sometimes blame themselves for unwanted sexual acts because of perceived failure to recognize contextual factors implying that sex is supposed to occur [62, p. 445]. Sexual scripts concerning consent are also gendered. For example, the gendered assumption that "men always want sex" can lead to the perception that asking for consent from men is "dumb" and unnecessary [54, p. 7].

Some HCI research has begun to consider the role that technology can play in shaping sexual consent practices. Wood and colleagues' mobile phone game *Talk About Sex* [116,117] intends to facilitate discussion about consent and other sexual wellness topics. Nguyen and Ruberg studied the "consent mechanics" in sex-themed video games to unpack "commendable" values around consent that are conveyed to players and which could be designed into other technologies [71].

Despite these positive examples, the literature indicates that technologies currently designed to augment progression to sex are perpetuating nonconsensual experiences. One example is mobile apps for recording consent between sexual partners, such as Good2Go and LegalFling. While intended to facilitate overt consent exchange, these apps have been harshly criticized in HCI literature [71] as well as popular media [76] for failing to design for the possibility that a person may change their mind about sex and want to stop the experience. LegalFling, for instance, turns consent into a legally binding contract [65], stripping users of any subsequent agency during the sexual act.

Online dating is another example of SV being inadvertently perpetuated through technology that is intended to augment sexual encounters. Several quantitative studies have linked dating app-use with SV [25,45,46,78,86,92,108]. Choi and colleagues found that dating app users are 2.13 times more likely to be sexually abused than non-users [25]. In two different studies of SV in Australia, dating apps were attributed to more than 10% of overall cases [78,86]. In a study of Tinder, users were more likely to report nonconsensual sex than non-users [92]. Other research indicates that dating app-facilitated SV is getting worse: reports of rape by perpetrators met through a dating app increased six-fold over a five-year period in the UK [108].

Our study explores computer-mediated consent in the context of online dating, partly because of the aforementioned research portraying it as an SV risk factor, and partly because of the widespread use of dating apps for pursuing sex (prompting some research to call them "sex apps" [13]). With 49% of US adults under the age of 30 having used a dating app [6] they are an opportune context for designing future interventions to mediate consent and therefore prevent sexual violence at scale.

## 2.3  Dating Apps: Sexual Opportunity and Risk at Scale

Dating apps are a subset of social matching systems [103], meaning that at a basic level their purpose is to introduce, or recommend, people to people. Examples include Tinder, Bumble, Grindr, OkCupid, and Hinge. Sex is a pervasive desire behind dating app-use [14,28,118,121], leading some users to think it is the predominant purpose of dating apps [26]. Some dating apps are more synonymous with sex than others, such as Grindr (an app for men seeking men) [15,39] and Tinder [36,98]. However, research indicates that motivations for using Tinder [102,104], Grindr [83,114], and dating apps in general [75,120] are more diverse than sex, including friendship, community building, validating self-worth, and long-term romance.

*2.3.1 Dating App Design.* Mechanisms for user discovery are rather uniform across dating apps today, with an emphasis on relative geographic location of users' mobile devices (e.g., "this user is 1 mile away"). Two features available to users for informing face-to-face meeting decisions are profile pages and private messaging interfaces [38]. Profiles are static representations of users that predominantly showcase pictures of their physical appearance. Non-picture content in profiles has witnessed a steady reduction since the transition from dating sites accessed on personal computers to dating apps on mobile devices. However it is still typical to find fields for demographic traits like age, open-ended text, and (in the case of apps for men seeking men) HIV status [111]. Messaging interfaces today commonly require two users to "like" each other's profile before messages can be exchanged, which is signified with an explicit notification of a "match." Dating apps also include various safety-minded features, including the abilities to block and report troublesome users. More recently Tinder has added a panic button that will alert trusted contacts if harm occurs during a face-to-face meeting [126], as well as a feature that maintains a record of users that one has met face-to-face.

*2.3.2 Use of Dating Apps.* Use of dating apps has usually been studied through the lens of impression management (self-presentation) [47] and signaling theory [32,33]. Impression management, originally from Goffman [47], has been used to study how online daters craft their profiles and message content to achieve an intended impression in the eyes of potential partners [43,113,120]. This is typically motivated by the desire to maximize attractiveness, which has driven users to engage in deception [38,43,48–50,105,106] and use profile and message content sold by dating coaches and recommended by strangers on the Internet [66,119]. The flip side of impression management is impression formation, or evaluation of other users to determine their value for future interaction. Aside from the challenge posed by deception, users have struggled with evaluating "experiential traits" such as personality [120,122], and have reported going on dates earlier than they wanted to in order to better collect this information [120].

Signaling theory describes the extent to which a piece of information indicates an otherwise unobservable trait [32]. A simple example would be a list of favorite books in a dating profile signaling intelligence [122]. Signaling theory has been used to study Grindr users' interpretations of profile content, such as interpreting use of PrEP (an HIV preventative medication) to signal desire for unprotected sex [111]. However users warn that signals of unobserved traits from profile and message content are susceptible to misinterpretation [122].

Ultimately, users have expressed frustration over their capacity to make informed face-to-face meeting decisions [41,120], which could lead to unnecessary exposure to physical risk. This has motivated prototypical interfaces such as the prompted discussion interface [123] and virtual dates [41] to better support impression formation.

*2.3.3 Risk in Dating Apps.* Risk has pervaded much of the online dating literature beyond the themes of deception and ill-informed face-to-face meeting decisions. Online dating poses amplified risk of HIV and STI transmission, particularly amongst users of Grindr [3,27,51,64,82,83,110,111,115]. Research has reported on users' hesitance with reporting HIV status in profiles: some users feel that disclosure may limit conversations around safe sex [110], while others purposely keep their profile scarce to maintain control over information that may be inadvertently disclosed [111]. Relatedly, there is evidence that users disguise, or blatantly lie about, their interest in casual sex due to fears of "slut shaming" and stigma [15,121], which can lead to misinterpretations of the reasons for meeting face-to-face.

Risks with self-disclosing other types of personal information have also been explored. Several studies have acknowledged the risk of being outed when using a dating app for men seeking men, which can have significant social and safety implications [12,15,27]. Transgender users also put themselves at risk when deciding to disclose their transgender status, although some users purposely disclose this information online so as to avoid potential harm in-person [37]. Users with disabilities have similarly self-disclosed online in order to weed out users with biases against those with disabilities [77].

Online harassment is another common theme in the online dating literature, which indicates women and LGBTQIA+ individuals being disproportionately victimized [2,6,26,93] by behaviors such as unsolicited pictures of genitalia, threatening messages, and sexually aggressive messages.

Safety in online dating has been a research focal point since at least 2001 [21]. That work has uncovered additional generalized safety strategies like avoiding users with no profile pictures [15] and increasing uncertainty reduction strategies [44], such as using search engines to investigate potential meeting partners. However the research indicates that dating apps are generally failing to protect users: Duguay points to suboptimal platform moderation [34] while Pruchniewska critiques Bumble for enforcing "gendered labor" by requiring women to carry the responsibility of vetting men [79].

Despite a history of studying risk and safety in online dating, sexual violence and consent practices have not been studied directly beyond quantitative associations [46]. There is some indication that this is a concern of users; Corriero reports on fears of rape and physical harm by users of Grindr [27] while Asbury and colleagues found that women are 3.6 times more likely than men to want to see consent-related information in dating apps [2].

## 3 RESEARCH QUESTIONS

We question how dating app design and user experience play a role in shaping and influencing users' processes of giving and (perceiving to) receive consent to sex, which may predispose them to engaging in or experiencing unwanted sexual activity. Our research questions are:

*RQ1:* *How do online daters exchange consent to activities of a sexual nature?*
    *RQ1A:* *How do online daters give consent to sexual activity?*
    *RQ1B:* *How do online daters identify consent to sexual activity from their partner?*
*RQ2:* *How do dating app interfaces and user experience inform users' consent processes?*
*RQ3:* *What challenges do users face with dating app-mediated consent to sexual activity?*
*RQ4:* *In what ways do users' dating app-mediated consent processes fail to protect against sexual violence (unwanted sexual activity)?*

## 4 METHOD

We conducted semi-structured interviews with 19 users of the dating app Tinder who represented a variety of gender identities and sexual orientations. The interview protocol involved participants giving detailed descriptions of sexual experiences facilitated through dating app-use from the point of initially discovering a user on the application to the point of penetrative sex occurring, or otherwise the final interaction with the user if penetrative sex did not occur.

Student researchers from demographics at elevated risk of sexual violence comprised most of the research team. They were motivated to address the prevalence of sexual violence and saw this research project as a vehicle for making a difference. These demographics included women, students identifying as LGBTQIA+, and members of Greek Life organizations. Greek Life—and its composites "fraternities" and "sororities"—refers to systems of social organizations at United States colleges and universities usually named with Greek letters [112]. They provide community service and postgraduation networking opportunities, but have a troubled history of conflict with college administrations and are associated with a culture of sexual objectification [61,85].

The student researchers co-constructed the interview protocol and recruitment methods. They also served as interviewers and co-led data analysis. Their research training, including IRB certification, and all other duties mentioned above were supervised by the paper's first author.

Table 1. Demographic details of interview participants.

| Referred to as | Gender Identity | Sexual Orientation | Ethnicity | Age |
|---|---|---|---|---|
| Kate | Cisgender woman | Bisexual/pansexual | White | 22 |
| Jose | Cisgender man | Heterosexual | Hispanic | 22 |
| Chrissy | Cisgender woman | Heterosexual | White | 22 |
| Willie | Non-binary trans-feminine | Pansexual | White | 27 |
| James | Cisgender man | Heterosexual | Asian | 25 |
| Billy | Cisgender man | Heterosexual | White | 20 |
| Alex | Non-binary trans-masculine | Bisexual/pansexual | White | 23 |
| Ethan | Cisgender man | Heterosexual | White | 24 |
| Grace | Cisgender woman | Heterosexual | White | 22 |
| Joe | Non-binary | Bisexual | White | 23 |
| Jess | Cisgender woman | Heterosexual | White | 20 |
| Chloe | Cisgender women | Heterosexual | White | 24 |
| David | Cisgender man | Heterosexual | Asian | 22 |
| Brick | Agender | Pansexual | White | 23 |
| Dan | Cisgender man | Pansexual | White | 23 |
| Tiglet | Cisgender man | Queer | White | 23 |
| Magmar | Trans woman | Bisexual | White | 24 |
| Bob | Cisgender man | Gay | White | 28 |
| Kyle | Cisgender man | Pansexual | Mixed race | 24 |

## 4.1 Recruitment

Recruitment methods involved student researchers promoting the study through social media posts, a word-of-mouth campaign that involved a text message invitation being shared amongst and between student social circles, paper flyers at social spaces (particularly LGBTQIA+-friendly spaces in which our student researchers had rapport), and snowball sampling. Recruitment targeted individuals in the 18-29 age range to map with statistics on dating app-use [6] as well as statistics on prevalence of sexual violence [95,96]. All recruitment materials emphasized that participants would be asked to describe their dating app-facilitated sexual experiences in detail. The recruitment approaches collectively yielded 19 interviews. See Table 1 on the previous page for demographic information (participants provided fake names for the purpose of anonymity). While recruitment materials invited users of all dating apps, Tinder was the common dating app used by all participants, and the primary app discussed during interviews.

## 4.2 A Note on Participant Care and Mandatory Reporting

Due to the nature of the study's topic, several precautions were taken to protect participants from re-traumatization and stress. At least one student researcher was present at every interview to establish comfort with the participant based on shared demographics around age and gender/sexual identity. All participants were asked before the interview began if they were comfortable with the interviewers in attendance and were reminded that they could end the interview or ask any interviewer to leave at any time. The research team also consulted with a practicing nurse with experience engaging with SV victims, as well as researchers with experience studying SV and related gendered risks, to understand best practices for participant inquiry. In order to understand how our responsibilities as mandatory reporters of Title IX violations extended to our study we consulted appropriate reporting structures as well as several faculty and staff at our university with experience conducting SV research and who were involved in implementing our university's mandatory reporting policy. It was determined that instances of SV reported through data collection were exempt from mandatory reporting; the decision was confirmed by our university's research office.

## 4.3 Data Collection and Analysis

All but two interviews occurred in-person at a location chosen by the participant, such as the authors' research lab, restaurants, and participants' homes. The other two occurred online. Interview lengths ranged from 27 to 62 minutes, with an average of 53 minutes.

After an IRB-approved consent form was signed, interviews began by assessing the participant's general online dating experience: which dating apps they used, how many online daters they had met face-to-face, their expectations for meeting those online daters, the typical outcomes of their face-to-face meetings, and their experiences with sex throughout the online dating process. Answering these broad questions naturally led participants to mention particular sex-related experiences they had through online dating. We encouraged participants to narrow their focus to these experiences and recollect them in detail, from the moment of initially discovering the respective user online up to the moment of penetrative sex or otherwise wherever their interaction with the user concluded. Questioning from interviewers served to clarify understanding of sexual interest and consent, particularly in the moments directly preceding vaginal or anal penetration, and other ambiguities. Once participants completed the retelling of their experience, interviewers probed additional experiences based on notes taken about particular moments in the online dating process (e.g., how a user's consent-seeking behavior compared to other sexual experiences they had through online dating).

All interviews were voice recorded and transcribed. Dedoose was used to facilitate iterative line-by-line coding of the transcripts according to Strauss and Corbin [99] and then recurrent review of the codes to refine emerging categories and themes. A team of three researchers (a pansexual non-binary person who was assigned male at birth, a heterosexual woman, and a heterosexual man) individually coded interviews and then collectively reviewed coded transcripts in recurrent meetings to refine the evolving codebook and motivate recoding of the interviews. Over multiple iterations all interviews were recoded to reflect the finalized codebook. The team then collectively engaged in axial coding of the transcripts to identify relationships between codes and organize them into a hierarchy. This hierarchy was used to identify selective/overarching code categories pertaining to computer-mediated consent processes.

## 5 FINDINGS

Selective codes during data analysis pertained to two distinct computer-mediated consent processes, which we call *consent signaling* and *affirmative consent*. These terms refer to the process adopted by a participant in a specific experience with a particular user for understanding, presenting, and receiving consent to sexual activity through Tinder's interface. These processes began at initial discovery of a user's profile page and ended at the point of physical sexual activity occurring (although in some instances interaction with the respective user ended without a physical sexual activity occurring). We unpack these two consent processes below, including the role that Tinder's interface plays in facilitating or influencing their adoption, and limitations of both consent processes that can leave users susceptible to sexual violence.

### 5.1 Computer-Mediated Consent Signaling

In experiences coded with *consent signaling* Tinder's interface was used to infer and imply consent to sexual activity without any verbal or overt confirmation before the activity was engaged in. These users (at the time that the respective experience occurred) understood Tinder to be an app intended to support discovery of sexual partners and rapid progression to sexual encounters. Profiles were thus perceived as tools for identifying nearby people interested in sex, rather than tools for evaluating *if* someone was on the dating app for sex. A "match" in Tinder's interface (when two users "swipe right" on each other's profile, enabling them to exchange messages) was considered a signal of mutual sexual interest and consent to sexually explicit messaging. Agreement to meet face-to-face through the messaging interface was then interpreted as agreement to have sex. Experiences coded with consent signaling were predominantly from users identifying as heterosexual men and women, and some LGBTQIA+ users when reporting on their early sexual experiences through Tinder with heterosexual partners.

*5.1.1 Profile Discovery Signals Interest in Sex.* Almost all participants coded with consent signaling assumed that the primary purpose of Tinder was for finding casual sex partners. An implication of this perception was that the mere presence of a profile page was interpreted as a signal of the profile owner's interest in casual sex. Tinder's profile discovery interface, in this light, was seen as a tool for effectively bypassing the need for overt discourse about sexual desire because it was perceived to identify who in the geographic vicinity wanted sex.

When participants reported how they evaluated the profiles of eventual or potential sexual partners, impression formation usually began and ended with determination of whether the evaluator was willing to have sex with the user presented in the profile. Profile pictures of physical appearance were the sole determinant of decisions to "swipe right" in most cases. As Jose described: *"There's like very few things that like, will get me to cause to swipe left. If you have, like, a really bad first photo, use too many, like, filters."*

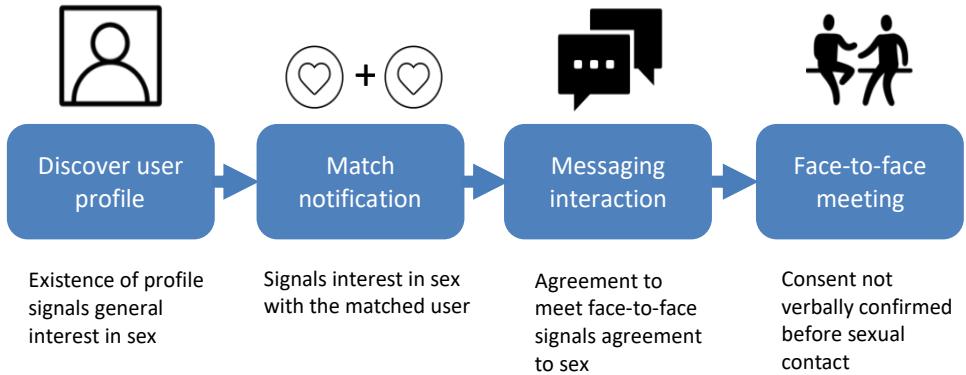

Fig. 1. Users practicing consent signaling interpreted the components of Tinder's interface as tools to effectively bypass the need for overt discourse around sexual interest and consent.

While Tinder profiles do have open-ended text bios, which can potentially hold information about one's interest in casual sex, participants usually ignored this content. In the few instances that bio content was mentioned participants dismissed its validity, especially if the bio content indicated that the profile owner was not interested in sex. Participants usually referred back to their preconception that Tinder is for sex when explaining why they found such content to be dubious for impression formation. Ultimately, the possibility that a profile owner may not be interested in sex was either rejected or not seriously considered.

**Ethan:** *"Some [profiles] are like, 'Oh, well, I'm here to look for a boyfriend' and I'm like, really? You're here on Tinder to look for a boyfriend? I can smell that bullshit a mile away. […]*

**Interviewer:** *"So those people that say things like 'I'm looking for a boyfriend' or things like that [in their profile]. Do you still attempt to, like, match with them? In pursuit of sex?"*

**Ethan:** *"Yeah, yeah. I mean, I always get, always try to give it the college try."*

Some participants explained that Tinder's interface design encourages sexual objectification of profiles. In Joe's words: *"Yeah, I would say that the swiping isn't, I don't think that's very good. The like/dislike feature based off of images alone. I feel like that just turns every user into a commodity or an object, you know, regardless of how they want to treat it or use it. It is a tool of objectification. So if you look at Tinder, you know, that is a tool of objectification for every user. Every user is objectified and objectifying others at the same time."*

Joe's quote represents a pattern of participants rationalizing the commodification of profiles as unavoidable by design. It does not simply turn them into commodities, but forces them to commodify others ("*every user is objectified and objectifying*") and therefore excuses them from any personal responsibility when they practice this objectification. Other participants pointed to the amount of screen real estate devoted to pictures and the necessity of an extra button click to read open-end bios as explanations for how interface design encourages them to evaluate profiles in this way.

*5.1.2 A "Match" in Tinder's Interface Signals Consent to Sexually Explicit Interaction.* Under consent signaling, "matches" in Tinder's interfaces (notifications of mutual "liking" between two users' profiles, which enables one-on-one messaging) were interpreted as signals of desire to have sex with the matched user and consent to immediately engage in sexually explicit interactions. Initial messages, which were always sent by men in our interviews, often immediately sexually objectified the recipient. These included overt statements of desire for sex, overt requests for sexual favors, or thinly veiled sexual innuendos. As James described: *"I would essentially be like, 'Hey, send me nudes,' or 'Hey, DTF.' Or hey, like, just really, really simple. Straight to the point. Like, 'Hey, you want to fuck?' Boom."*

These messages were the first attempt in the consent signaling process to overtly state interest in sex with another user. However, the intent of these messages was not to clarify mutual sexual desire, but to expedite the progression to a face-to-face encounter so that sexual urges originally developed during profile evaluation could be acted on. Men exhibited some frustration when partners were not receptive to these messages. For example, David was annoyed with "*one word responses*" to his messages soliciting sex: *"Yeah, yeah, like one word responses. It's the most annoying thing. Like I come up with like a witty three liner and you come with 'KK All right.' I'm out. I'm done. Like you, you just wasted my time. Yeah, like, I don't want to meet you then."* This frustration with "wasted time" was reflective of the perceived purpose of the messaging interface being to escalate online interaction to a physical sexual encounter. Refusal by a messaging partner to abide with sexual objectification was therefore considered an incorrect use of the application.

Recipients of sexually objectifying messages concurred that these types of messages were normal and that use of Tinder for non-sexual reasons falls outside of its intended use. Importantly, they maintained this perception even when they themselves used Tinder for reasons other than, or in addition to, sex (their choice of Tinder in those cases was driven by its sheer popularity or because the participant's social circle used the app). This occasionally impacted their understanding of sexual agency, or their capacity to reject sexual requests over messaging. In one instance Grace described how she was hesitant to firmly deny a man's request for *"full body"* pictures because the request aligned with her understanding of how Tinder's messaging interface is supposed to be used. As she described the interaction:

**Grace:** *"So I remember he said can you send me a picture because your ones on [your profile], they were old or whatever. So I sent him like a face shot. I'm in the library. So I can't really do much, right? I just take a selfie. [He was] like, no, I want a full body shot. So I was like, okay, whatever. It's fine. I'll just go into the bathroom. Took a full body selfie in the mirror. He's like, well, I was hoping for more than that. [...] I told him no, indirectly. So it was kind of like, I'm not. I don't know. I was like, teasing and flirting back but like saying no. You know what I mean?"*

Grace went on to rationalize the man's behavior as understandable because of his sexual drive and therefore undeserving of a stern rejection. She described this kind of messaging exchange as typical of her Tinder experience. In her words: *"They're horny. And some of them are rude. I've had negative experiences with most of the guys on there. They, I mean, I guess not negative in their sense. Like he was there to hook up."* Chrissy similarly considered sexual solicitations over messaging to be a normal and acceptable experience, despite such messages driving her to stop using Tinder. Rather than voice her disapproval she would apologize to men for her lack of interest in casual sex because she considered her reasons for using Tinder to deviate from the supposed sexual purpose of the app.

**Interviewer:** *"Why'd you stop using Tinder?"*

**Chrissy:** *"I didn't, I didn't like the conversations. I didn't find anyone that was like, good enough to talk to or else the conversations did get weird [...] in a sexual way where they're just like, well, 'I'm* *just here to fuck.' I'm like, well, 'I'm not. Sorry.' [...] I mean, it is Tinder. So like you're expecting it at some point. [...] When it happens you're not surprised. Like, oh, well this is Tinder."*

In only one experience of consent signaling did a recipient of a sexually objectifying message (Chloe) describe firmly rebuffing the advance. She explained how men would often react with confusion because they understood the purpose of messaging interaction to be for arranging a sexual encounter: *"a lot of times they'll be like, 'why else are we here?' Like on this app, like a lot of people just think Tinder is a hookup app."*

*5.1.3 Agreement to Meet Face-to-Face Signals Consent to Physical Sexual Contact.* In experiences coded with consent signaling, participants often talked about the prospect of face-to-face meeting and sex interchangeably, as if sex was the assumed purpose and expectation of face-to-face meetings. While men's intent to meet for sex was usually clear through their sexually explicit messaging, mutual intent from users who were not cisgender men was typically not as clear. When describing how they *"knew"* a messaging partner wanted to meet and have sex with them, men pointed to the content of messages received from their partners, however examples never involved the messaging partner confirming or clearly stating that they wanted to have sex. Rather, desire to meet face-to-face for sex was interpreted from message content such as jokes of a sexual nature, physically revealing pictures sent through Snapchat or text messaging, or even just the absence of an attempt to change the messaging topic to something less sexual. James described such responses as *"reciprocating the energy."* And in Ethan's words: *"She was receptive [to sex]. I'm assuming [that from] what kinds of things she was saying."*

Men sometimes described particular moments in messaging interaction that solidified their interpretation that a user was interested in meeting for the purpose of sex. Billy exemplified such a moment with an exchange of sexually explicit pictures, which led him to invite a woman to his house at the next opportunity that his parents were not home.

**Interviewer:** *"How did that change your perception of things, once you exchange those [sexually explicit] pictures?"*

**Billy:** *"I got a little more hopeful that she was going to come over and we're going to end up hooking up."*

A consequence of assuming sex to be the purpose for face-to-face meetings was that participants sometimes conflated agreement to meet face-to-face with agreement to sexual activity. In one example, Ethan described how his messaging interaction with a woman solidified his interpretation that she wanted to have sex with him. This was based on her confirmation that she was one time a *"Catholic school girl"* which he interpreted as a signal of interest in anal intercourse. Sex became not just the assumed purpose of their face-to-face meeting, but something Ethan felt entitled to as a condition of spending time to drive to the woman's town.

**Ethan:** *"[During messaging] she was like, 'Yeah, I was a Catholic school girl.' She gave me a winky face [emoji]. It's like, alright. Yeah. I know. I know what she's trying to do here. [...] I was really hoping for sex. Which was a lot, which is, which, which like, okay, here: a guy doesn't drive over an hour and leave empty handed."*

Ethan later described, with laughter, the anal sex that occurred during his face-to-face meeting with this woman as fulfillment of the Catholic school girl fantasy originally developed in their messaging interaction: *"I'm laughing because I'm remembering how funny [the anal sex] was because the stereotype of Catholic school girls being very, very, very, very, very kinky."*

Like Billy's excerpt above, Ethan mentions *"hoping for sex,"* which suggests that men's interpretations of message content were biased by their own desires for sex (a confirmation bias).

However, in experiences coded with consent signaling men seemed oblivious to the subjective nature of their interpretations of agreements to meet face-to-face. None of them made attempts to confirm their interpretations—either online or in-person—that the purpose of meeting face-to-face was for sex.

This can create dangerous situations when such interpretations are wrong. Chrissy demonstrated this with a story about a man following her to a bar. He knew she was spending time at a particular bar because she had mentioned it in a message, which he interpreted as an invitation. He never confirmed this interpretation with Chrissy before showing up to the bar, at which point she used her friends to hide from the man.

**Chrissy:** *"[I told him] 'I'm going to karaoke night to the bar by my house.' I didn't really expect it to be an invitation, but he took it as one and showed up and I was there with my friend it was really awkward. [...] I was, I literally, I was sitting at the bar and watched [him] walk up with his friend I was like someone hide me, like please I swear to God, I swear to God, and then he walked in and I was with my roommate at the time, and her friend and then my neighbor, and we were all kind of sitting at the bar. And the dude comes in and I stole my [...] neighbor's seat. I was trying to like hide from him."*

## 5.2 Risk of Sexual Violence with Computer-Mediated Consent Signaling

In no experience of consent signaling was consent verbally confirmed prior to a physical sexual activity occurring. This poses a risk of nonconsensual sexual activity because of the possibility that signals of consent could be misinterpreted. The initiators of physical sexual contact through consent signaling—all cisgender men—also conveyed problematic conceptualizations around consent. For one, they believed that their partners did not want to overtly discuss consent. In Billy's words: *"I know most girls don't want to, like, talk about that kind of stuff. Unless they're the ones bringing it up."* Due to the assumption that sex was already agreed to upon meeting face-to-face, other participants understood consent to be confirmation of *"when"* sex would occur, not if it was desired. Similar to Billy, James considered consent discussions to be disrespectful to women: *"I don't think being super open about it is like, the best approach. Like I don't really, I'm not really going to be like, hey, when are we gonna fuck? Like, you know, I don't feel like being, that's like, I don't feel like that's very respectful."*

In lieu of verbal discussion of consent, cisgender men believed that they could *"sense"* when their partner was ready to have sex. This was typically signaled through a lack of physical resistance to a sexual advance. For example, Jose assumed a woman from Tinder wanted to have sex because she did not overtly refuse his attempt to pull her shirt off: *"It's like a sense thing. I can't really describe it. [...] I grabbed her shirt and I was like, pulling it off. And I guess she was just okay with it. [...] It just happened. Like, there was no like verbal like, 'hey, do you want to do this?'"*

Other participants described letting a sexual experience happen that they otherwise did not want for reasons including a perceived obligation to have sex and fear of physical retaliation. For example, Tiglet described feeling a *"need to, like, perform"* while in a woman's bedroom that he met from Tinder. He did not want to have sex and was unable to have an erection, but he did not physically resist when the woman placed his hand on her breast and attempted to stimulate herself.

**Tiglet:** *"It's important to get, you know, consent from people in this day and age, but I also don't think we're at a point where that always happens explicitly, verbally, at least not for the first time. I mean, it's good if that happens. But you know, we're in [the] bedroom, doors closed. [...] The first time*

*I touched her boob she put it there when I think about it. [...] I was unable to get hard and she did whatever. Maybe it was still a feeling and need to, like, perform."*

Chloe described a different experience where she let a man from Tinder kiss her while she was sitting in his truck, despite not wanting the experience, because she feared retaliation if she resisted. As Chloe described: *"I was afraid. Because I was in his car, like it's his environment. He could've just driven off. Like lots of times we're afraid if we don't give him what he wants something bad will happen. I'd rather kiss someone than, you know, die."*

### 5.3 Computer-Mediated Affirmative Consent

In experiences coded with *affirmative consent* Tinder's interface was used to prompt overt discourse around consent before face-to-face meetings. The term *affirmative consent* comes from public health literature [58] and legislation [23] that rejects the lack of resistance to sex as a form of consent and necessitates that partners give unambiguous agreement to sex ("yes means yes"). Tinder's interface was conceptualized by some participants as affirmative consent software, enabling them to foster patterns of affirmative consent and identify resistance to affirmative consent practices in a reduced-risk computer-mediated environment. Consent to specific sexual acts would be overtly exchanged before ever meeting a sexual partner face-to-face, and then verbally reconfirmed during the face-to-face meeting. Experiences coded with affirmative consent were entirely from users identifying as LGBTQIA+ and were typically adopted in response to sexual violence or general harm that such users experienced in their prior dating app-use.

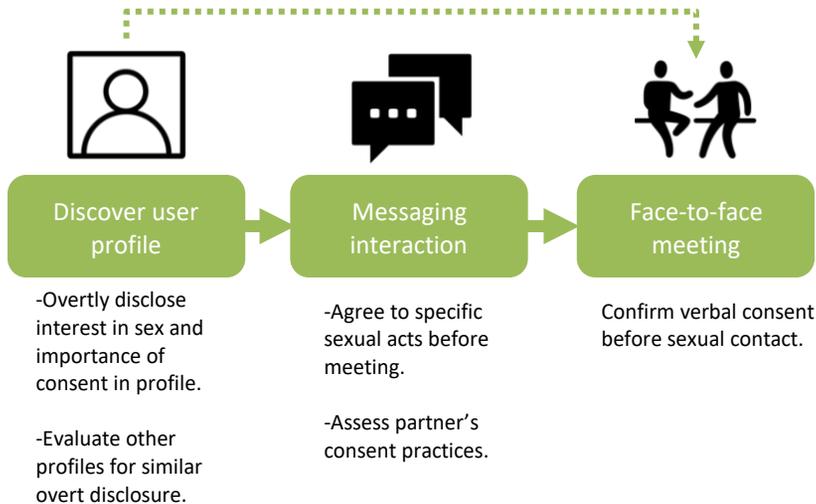

Fig. 2. Users practicing affirmative consent used the components of Tinder's interface as tools to foster overt discourse around consent and identify resistance to affirmative consent practices.

*5.3.1 Profiles are Tools for Overt Disclosure of Interest in Sex and the Importance of Consent.* Adoption of computer-mediated affirmative consent was commonly driven by prior experiences of *"sexual trauma"* (Alex) that involved the participant personally being sexually assaulted or a previous sexual partner disclosing trauma. Willie described a sexual experience that motivated their adoption of affirmative consent in this way: *"One time I was hooking up with someone and they started like crying halfway through and that really just, for lack of a better word, fucked me up.*

*I was like 22. And apparently they had been assaulted."* Adoption of Tinder as affirmative consent software was intended to mitigate future experiences like this for participants and their sexual partners.

One way that participants did this was by crafting the open-ended text bios in their profiles to clearly state openness to sex, amongst other social goals, so as to mitigate any misinterpretation for why they were using the app. This profile content was often the result of an iterative, trial-and-error process. As Kate described: *"I think I re-worded my bio so that I, I was open to hooking up. [...] I've gotten it to be pretty clear about, like, what the expectations were."*

Several participants described how these iterative profile revisions led them to include open-ended content about consent itself, particularly the importance of exchanging consent during sexual encounters. This reflected their own personal valuing of consent as well as the role of overt consent disclosure in their preferred process of escalating to a sexual encounter.

**Dan:** *"For a very long time on my profile, I did have, you know, that near the very top of my profile, especially when I was looking for casual sex there, I did put consent is extremely important to me. Like, if you're not willing to have conversations even about consent, like please do not expect anything like that. [...] I definitely put it there just because I value those things."*

**Willie:** *"I think it just literally is like, I'm very straightforward. 'Looking for hookups or friends with benefits, if consent is discussed at length.' I think that is my bio verbatim right now."*

When evaluating the profiles of other users discovered in the app, participants similarly looked for open-ended bio content that overtly disclosed interest in sex and valuing of consent. This became particularly important for participants looking for relatively uncommon sexual arrangements beyond one-time sexual encounters such as non-monogamous or polyamorous relationships. Dan recounted a positive experience with a Tinder user who included similar profile content about non-monogamy: *"I said that I'm looking for friends, yeah, I'm looking for casual hookups, I'm looking for relationships. And I mean, within my profile, I put 'I practice non-monogamy' and she had a pretty similar setup in the bottom of her profile too."*

Participants found that other LGBTQIA+ users, as well as users that self-identified with the "kink" community, were most likely to include profile content about sexual desire and consent. They indicated that cisgender straight men were the least likely to include this content, and a few participants indicated that they completely avoid cisgender straight men for this reason.

*5.3.2 Messaging Interaction is a Tool for Confirming Compatibility of Expectations.* In a few instances participants only used profiles for mediating their affirmative consent processes and would not overtly discuss sex and consent again until the face-to-face meeting. As Kyle described: *"I try and save a lot of the conversation for the in -person thing, because otherwise, if you get all your messaging done right away, it's going to be a very awkward first date."* However, in a majority of experiences coded with affirmative consent, messaging interaction through Tinder's messaging interface as well as external messaging tools such as Snapchat and SMS played a significant role in mediating consent.

A common topic of messaging interaction after two users "matched" was to reiterate or further explore each other's goals for using Tinder. Rather than take the form of sexually explicit and objectifying interaction as was previously mentioned in the consent signaling findings (5.1.2), sex was broached in these messaging conversations in order to confirm that one's messaging partner was comfortable with sex and understood that they had the option to end the interaction if they were not. As Joe described: *"Well, for casual sex, it was mostly I was like, pretty forward that I want everybody to be comfortable with what's going on. And so if that's not what you're looking for, I don't want to push that."* Several participants practicing affirmative consent were using Tinder for a variety of goals ranging from friendship to casual sex and long-term relationships, and so it was also important to them to clarify which of their many goals a messaging partner aligned with. This became most important when the participant was not interested in sex with their messaging partner and wanted to clarify that misalignment before exposing themselves to sexual risk face-to-face.

**Dan** (recounting a situation where he did not want sex, but his messaging partner did): *"I pretty quickly was just like, hey, you know, I'm not here for anything like that. I'm in a happy relationship, but I'm not really looking to expand or do anything else right now. I appreciate it. I'm flattered. And I'm down to talk about these sorts of things with you, but I just want to, you know, make my intentions clear. I'd like to sit down, have coffee and talk about what it means to be an undergrad in PoliSci, as opposed to do that and then we bang my car."*

Once participants confirmed compatibility of interest in sex, messaging interaction evolved into overtly discussing sexual boundaries or the specific sexual activities that both partners wanted to engage in. As Willie described a recent experience: *"I think the first thing we talked about, we, we wanted to give each other oral first, and they volunteered to go first."*

These conversations were also an opportunity for participants to clarify sexual acts that they were not comfortable with and would want to avoid during face-to-face meetings. In Tiglet's words: *"my interactions there were more like, you know, I'd be honest and forthright like, hey, I've never done anything with a guy but I'm interested in it. I might be down to do like, do the blow each other stuff, but like I'm probably not going to want to do anal right now or anything like that."* Relatedly, conversations tended to also explore sexual history, particularly concerning HIV, STIs, and users' history of getting tested.

A few participants indicated how their profile content worked in conjunction with messaging to facilitate conversations about sexual boundaries. Matched users would utilize participants' profile content about consent to voluntarily broach discussion of sexual boundaries, which alleviated the need for our participants to have to navigate towards these topics. Two participants indicated this was more likely to happen with messaging partners from the "kink" community (users who disclosed specific and relatively niche sexual desires in their profile).

**Dan:** *"I would definitely say that a noticeable amount of the time, roughly half of the time that I'm talking about things regarding sex for consent people, it comes about through things that were previously talked about in my profile."*

Trust and comfort were recurrent motivations in participants' descriptions of their messaging conversations concerning sex. The intent of sexual boundary discussions was not to build anticipation of sex or to increase the odds of sex occurring. Beyond the surface level conversation topic of planning the sexual activity, participants intended the conversations to build trust and make their partners feel comfortable with maintaining overt dialogue about willingness *and hesitance* to engage in sex. The messaging interface was a tool for reaffirming, rather than abandoning, sexual agency. While this posed ample opportunity for one's messaging partner to back out of a sexual encounter, participants believed it made subsequent sexual encounters "better" because of the trust and comfort underlying them.

**Joe:** *"There's a lot that can go into it that makes it way more fun, interesting, engaging. That sort of thing. So, I would want to have a fairly, fairly thorough discussion about likes and dislikes, fears and desires in sex so that way, you know, if at any point you or the other person wants to withdraw their consent, they can feel comfortable in doing so because I made it readily apparent that I want to respect their boundaries. So hopefully that person feels more trust in me. I'm making the sex better."*

*5.3.3 Messaging Interaction is an Opportunity to Test Consent Practices.* Like profile evaluation, participants utilized messaging interaction through Tinder and other messaging apps such as Snapchat to detect signals of resistance to affirmative consent and thus risk of sexual violence during potential face-to-face meetings. One signal was a general unwillingness to discuss sex in detail over messaging conversation. In Kyle's words: *"That's a pretty big red flag if someone's not willing to openly discuss what the expectations are, because then you're probably going to be taken advantage of, or there's just something, something's up."*

Another strategy was using Snapchat to intentionally make oneself susceptible to online sexual harassment. Snapchat is a separate social media application that affords users the capability of exchanging ephemeral picture content (including sexually explicit photos), which some participants saw as an opportunity to evaluate other users' practices around consent. For example, Kate described how she would intentionally use Snapchat to give messaging partners the opportunity to send sexually explicit photos without permission. If a messaging partner did send such pictures without asking her first she would interpret that as a signal that affirmative consent practices may not be respected and followed during a face-to-face encounter.

**Kate:** *"I usually don't meet people [right away]. If I'm just messaging them on Tinder, usually give them like social media first, and then if they don't send me like, pics there, genitals, like, then it's like, a good sign. […] With Snapchat, it's pretty easy to like, snap a picture of your genitals and send it to somebody. So I mean, if they're not doing that, without my expressed permission, then I think that speaks to them being a decent human being."*

*5.3.4 Consent is Overtly Exchanged Online and Reconfirmed In-Person.* Discussion of sexual boundaries over messaging naturally led to overt agreement to specific sexual activities before face-to-face meeting. Upon meeting face-to-face participants described reconfirming consent verbally before engaging in physical sexual activity. Kate described this as consent *"along the way."* Participants gave two reasons for reconfirming consent in-person. One, it gives each partner an opportunity to change their mind. This possibility is particularly high when using dating apps because of the potential time lapse between messaging interaction and face-to-face meeting, and new information that users can gather during face-to-face meetings that corrects or supersedes impressions formed online. Some participants exemplified this with profile pictures being deemed physically inaccurate during face-to-face meetings, which reduced sexual desire. Others spoke of unanticipated anxiety upon meeting face-to-face.

Another reason pertains to perceived pressure to engage in sex. Several participants spoke of feeling an expectation to perform sexually, especially when intent to have sex was already explicitly stated over messaging. Dan elaborated by describing how he was less willing to decline any sexual advance when he was younger, regardless if he really wanted the experience, because he was *"desperate"* and used sex to affirm his sense of self-attractiveness. Reconfirming verbal consent in-person was thus an opportunity for his sexual partners, who may similarly view sex as a vehicle for self-worth, to reflect on whether they really wanted the experience to happen.

**Dan:** *"I believe in consistent, enthusiastic consent. So you need to enthusiastically and consistently over a period of time affirm with me that sex is something that you want to have for me to genuinely believe because I remember what it was like being a desperate person on dating apps. I had OkCupid when I was that weird and dirty ugly duckling. I know that people will be like, yeah, yeah, whatever. Yeah, let's do it. Let's do it. I just really want to [confirm that]."*

## 5.4  Risk of Sexual Violence with Computer-Mediated Affirmative Consent

Despite computer-mediated affirmative consent being adopted by users specifically to mitigate nonconsensual sex, participants exhibited reasons why this process can still expose users to sexual violence. For one, use of the dating app for scaffolding affirmative consent stops once users transition to face-to-face encounters. Once-overt dialogue about sexual boundaries and consent over asynchronous messaging can make way to ambiguous face-to-face situations where the need to (re-)exchange consent becomes *"blurry"* (Willie), leaving victims to hesitate in moments of nonconsensual contact. Willie recounted such an experience of being raped by a Tinder date. After they had a consensual sexual experience their partner attempted to initiate penetrative intercourse for a second and third time but without renewing consent. Willie doubted their sexual agency in that scenario and hesitated to stop the nonconsensual experience.

    **Willie:** *"[I was thinking] I hope they don't want to do this again. And they sort of grabbed me and like, started having sex with me. So looking back on it, I could definitely say they assaulted me. It's just like, consent gets really blurry. And I should, just I don't know, that's just what happened. I didn't really like fully realize a lot of this until like a couple years later, thinking about bad experiences."*

Willie connected this to another complicating factor that may obfuscate consent face-to-face: alcohol consumption. The literature indicates that people under the influence of alcohol are unable to freely give consent [11], which may negate the validity of consent exchanged online before meeting, and obfuscate whether consent is—or even can be—exchanged face-to-face. Other participants remarked about general awkwardness with broaching conversations about verbal consent during face-to-face meetings without the use of an intermediary tool such as Tinder's interface. For example, Kate discussed struggling to get a sexual partner to verbalize unambiguous consent during a face-to-face meeting because it clashed with the partner's preference to feign reluctance to sex that they really wanted.

    **Kate:** *"One time that I was like, really confused, because this was like, the first time I hooked up with like, a woman. And I was trying to, like, get her consent for things. And she like, said some, she said stuff that she didn't mean, like, she'd be like, 'no, no, no, no,' like that. […] And I was like, 'are you, like, are you good? Do not want me to do this?' And she'd be like, 'no, no, yes!'"*

Participants reported additional struggles with trying to implement computer-mediated affirmative consent through the dating app. The strategy of overt discourse around sexual boundaries and consent was reported as uncommon amongst the broader userbase by all participants practicing affirmative consent, which had implications on how the strategy was perceived by others. While disclosing sexual desire in their profiles sometimes helped trigger conversations about sexual boundaries, other times it was misinterpreted as an invitation for sexual objectification. Several participants remarked that they endured frequent sexually objectifying messages as a result of being open about their interest in sex. Likewise, participants acknowledged that their messaging partners often misunderstood their attempts to discuss sex as objectification. As Joe described:

    **Interviewer:** *"So did you find that other users seem to express that same desire and wanting to talk about consent [over messaging]?*

    **Joe:** *"I'd say if anything more feminine people were willing to do that then masculine people, but most people thought it was weird. Most people thought that it […] felt more objectifying. […] You're still assuming that that person is going to be okay with even getting that type of message."*

Similarly, other participants reported how messaging partners sometimes misunderstood attempts to discuss consent as *"sexting"* (Kate) or invitations to have erotic conversations with the intent of building sexual desire rather than navigating sexual boundaries. In Kate's words: "*I think some people think that if I'm talking about sexual stuff. If I'm talking about those subjects [like consent], it's automatically like, dirty, or like, a sexual scenario.*"

## 6 LIMITATIONS

The study's focus on Tinder may hamper generalizability of the findings to other dating apps. This is likely for dating apps with different design structures, such as Grindr which does not include a swiping/"match" mechanism, and Bumble's mandate that women send the first message after a "match" occurs. While the age range of our participants aligns with statistics on SV victimization [95,96], our sample leaves questions regarding dating app-mediated consent in other age ranges. Our sample is also predominantly white and located in the Midwest United States; the findings may not reflect the experience of users from other ethnicities or locations. Due to the cross-sectional nature of our study design, it is also impossible to know how the process of using Tinder over time may have gradually shaped our participants' understanding of consent or their modes of consent exchange. As was particularly the case with the users adopting affirmative consent, computer-mediated consent processes can be iteratively refined, and our interviews likely disproportionately reflected participants' most recent consent processes.

## 7 DISCUSSION

We conducted a semi-structured interview study with 19 users of the dating app Tinder to explore how processes of exchanging consent to sex are computer-mediated. Two computer-mediated processes of consent exchange were discovered: 1) consent signaling, in which Tinder's interface was used to collect and convey escalating signals of sexual interest and consent, but without any explicit confirmation of consent prior to initiating physical sexual contact; and 2) affirmative consent, in which Tinder's interface was used to establish patterns of overt discourse around sex and consent to specific sexual activities, which was verbally re-confirmed upon meeting face-to-face. Limitations of both computer-mediated consent processes were discovered that can result in nonconsensual sexual activity occurring.

In this section we first use our findings to pose explanations for the frequency of online dating-facilitated sexual violence (SV), and reflect on how the discovered computer-mediated consent practices compare to offline consent practices. We then discuss how Tinder is adopted by users not merely as a dating app but as a consent exchange app, and why dating apps need to more deliberately design for consent scaffolding in order to accommodate this latent motivation behind app-use. We conclude with suggestions for intentionally designing consent mediation, and the potential of dating apps to become scalable solutions for SV prevention.

### 7.1 Explanations for Why Online Dating is a Sexual Violence Risk Factor

There is mounting quantitative evidence that online dating is contributing to sexual violence (SV), meaning sexual contact without mutual consent [25,45,78,86,92,108]. However there has been a lack of empirical insight into why online dating perpetuates SV [46]. Our study poses two explanations. One is that the process of using Tinder leads users to assume that consent to sex has already been given by virtue of online interaction through the app and agreement to meet face-to-face. Two, the process of using Tinder can obfuscate one's sense of sexual agency—the perceived ability to decline a sexual advance [84].

The media psychology and trauma literature [7,53] has warned that the Internet is a powerful vehicle for mass modeling of behavioral patterns including sexual scripts [94], or socially learned perceptions of appropriate sexual behavior. Sexual scripts are at the root of many problematic consent practices and explanations for SV in offline contexts [22,40,54,62]. Unspoken meanings behind contextual factors like place and time (e.g., being in a college dorm room late at night) and gender (e.g., "men always want sex") [62] lead individuals to assume consent has already been given, or cannot be revoked. Our findings show that Tinder not only enables "mass modeling" of conventionally harmful sexual scripts, but that new sexual scripts unique to Tinder-use have emerged. Tinder's sexual scripts carry expectations of sexual interaction much in the same way that physical spaces and contexts do [22,54]. They also carry similar behavioral outcomes: assuming consent and doubting one's ability to say no.

The obfuscation of sexual agency through Tinder's sexual scripts makes apparent why existing safety features in Tinder, and dating apps more broadly, are ineffective at stopping online dating-facilitated SV. Features such as user blocking, reporting, and the more recent panic button [126] all necessitate that victims recognize when a "wrong" behavior is occurring. That recognition may never come if SV-qualifying behavior is interpreted as normal, acceptable, and aligning with the sexual scripts of Tinder-use. Regarding SV statistics, this suggests that dating app-facilitated SV is likely underreported.

### 7.2 Tinder…The Consent App?

HCI researchers have exhibited a growing interest in mediating consent to sex through technology [71,116,117], but have been critical of existing solutions [71]. Our study shows that Tinder is used not simply for discovering sexual partners, but also for scaffolding the consent process, suggesting that researchers should be considering dating apps as additional consent exchange technologies.

Users practicing consent signaling interpreted Tinder as a tool that expedites consent exchange through bypassing the need for it. One's presence on Tinder, and subsequent interaction through the interface, provided signals of sexual interest—in other words, users thought it performed consent exchange on their behalf. Users practicing affirmative consent, on the other hand, used Tinder because it afforded them control over the consent process; it enabled them to impose their preferred process of overt consent discussion in a reduced-risk computer-mediated environment.

Under neither interpretation is Tinder a particularly good consent exchange app. Under consent signaling, consent is implied through information that is not directly about consent, and not directly about sex at all in most cases, such as a "match" in Tinder's interface or an agreement to meet face-to-face. As reported on in prior HCI literature regarding signaling theory, signals can vary in their reliability [32,33,122]. It would seem to the impartial observer that signals of consent and sexual interest through Tinder are extremely unreliable.

Computer-mediated affirmative consent, while less susceptible to misinterpretation, also has limitations. For one, the role Tinder plays as a consent exchange app ends prematurely. It does not facilitate consent exchange during face-to-face meetings where risk of SV is at its highest, and users in our study recounted struggles with reconfirming consent or refusing sexual advances in-person even after establishing open dialogue through the app. In addition, unlike consent signaling, participants acknowledged that affirmative consent is an unusual use of Tinder's interface, and it was a struggle to stay committed to the process while enduring sexual objectification and negative reactions.

## 7.3 Computer-Mediated Consent by Design

The ways in which Tinder mediates consent are currently unintentional through design. Dating apps should attempt to deliberately scaffold consent because, otherwise, misalignments in consent processes and misassumptions around receiving consent stand to continue the perpetuation of SV through online dating. Yet it is shortsighted to think that mediating consent through dating app design would only mitigate SV *facilitated* by online dating. It would reframe dating apps as scalable, generalized tools for SV prevention given the significant proportion of adults that use dating apps [6].

Before intentional design for consent mediation can happen, designers must decide which consent exchange process they want to support. Affirmative consent is the practice currently advocated in the literature, law, and by sexual health organizations [23,58,91,124,125]. Affirmative consent requires mutual, overt agreement from all partners to the initiation and escalation of sexual activities ("yes means yes"). Because partners must actively give and seek unambiguous agreement recurrently as sexual activity escalates, it minimizes opportunity for unwanted sexual activity due to misinterpretation of desire or misunderstanding of one's own right to stop a sexual act. However, affirmative consent is not without its challenges [58]. The practice is not commonly followed and can seem awkward or mood-killing during real world situations. Our own study showed that affirmative consent can be too sex-forward for the masses. As some participants pointed out, there are users who simply do not want to discuss sex. This is poignant given that dating apps are now used for goals beyond dating and sex, such as friendship and employment [63,104,120,127]. It would be presumptuous of users' goals to mediate consent with features that scaffold overt discussions about sex in particular.

One possibility, per Nguyen and Ruberg [71], is to broaden the concept of consent beyond sex. Dating app interfaces could be designed to make overt discourse around consent for all kinds of social interactions more natural and normative. The swiping feature now typical in mobile dating apps for exchanging consent to messaging interaction is one example, and could be replicated for other escalations in interaction. Users could "swipe" to explicitly indicate a willingness to meet face-to-face, and for what reasons (e.g., dating, friendship, employment).

Another opportunity for design would be to consider features to be used during face-to-face meetings. Some participants in our study exhibited frustration with maintaining overt discourse around consent upon meeting face-to-face. Without the dating app to assist them users had trouble broaching conversations about consent, vocalizing disapproval to a sexual advance, and recurrently receiving verbal consent to escalating sexual activities. When considering ways to mediate in-person consent exchange, critique of consent exchange apps like LegalFling and Good2Go [71] highlight the risk of inadvertently stifling sexual agency if users cannot revoke computer-mediated consent. Potential approaches for encouraging overt and recurrent consent exchange that provides opportunity for users to revoke consent could be app-recorded consent that "times out" after a certain period (therefore requiring re-exchange) or a conversational UI that requests verbal consent from partners when the application has not sensed any recent verbal exchange.

Above all, we think the best course of action for future work is to follow the principles of feminist HCI [9], specifically participation. Artifacts designed by researchers who may not personally use them hold the potential to inadvertently obstruct sexual agency in ways similar to existing consent exchange apps. Involving end-users in the design of consent-mediation tools would ensure that consent exchange is scaffolded in ways that protect their sexual agency and accommodate factors that researchers and designers may not have anticipated.

## 8 CONCLUSION

This paper reports an interview study about computer-mediated consent to sex in the context of the dating app Tinder. The study found that users conceptualize Tinder as a consent exchange app, albeit in different ways. Users practicing *consent signaling* considered Tinder to effectively bypass the need to overtly exchange consent to sex because one's presence on the app, and any subsequent interactions through the app, were assumed as signals of sexual interest. This poses risk of sexual violence (nonconsensual sexual activity) because such signals could be misinterpreted. Users practicing *affirmative consent* used Tinder's interface to scaffold overt discourse around sex and consent before meeting face-to-face so as to minimize threat of sexual violence. However, such users were still at risk of sexual violence because they found it difficult to maintain overt consent exchange in-person where the app could no longer be used to scaffold sexual dialogue. Ultimately, the paper argues that dating apps could become valuable tools for computer-mediated sexual wellness if designed to intentionally mediate consent exchange.

## ACKNOWLEDGEMENTS

We thank Chris Aiello, Daniel Yu, Marcel Jonas, and Steven Said Jr. for their contributions to participant recruitment and data collection. We also thank Shelby Pitts for contributions to data analysis. The constructive feedback from anonymous reviewers of this paper is also much appreciated. This research is supported by a Faculty Research Fellowship from the University Research Committee at Oakland University.